\documentclass[aps,twocolumn,prl,floats,superscriptaddress,amsmath,amssymb,showpacs]{revtex4-1}

\usepackage{epsfig}
\usepackage{graphicx}
\usepackage{color}
\usepackage{mathrsfs}

\def \kstar{{K^{\!*}}}

\def\beq{\begin{equation}}        
\def\eeq{\end{equation}}
\def\bea{\begin{eqnarray}}
\def\eea{\end{eqnarray}}
\def\Im{{\rm Im}}
\def\Re{{\rm Re}}
\def\nn{\nonumber}
\def\sss{\scriptscriptstyle}

\def\barp{{\raise.35ex\hbox{${\sss (}$}}---{\raise.35ex\hbox{${\sss )}$}}}
\def\bdbarp{\hbox{$B_d$\kern-1.4em\raise1.4ex\hbox{\barp}}}
\def\bsbarp{\hbox{$B_s$\kern-1.4em\raise1.4ex\hbox{\barp}}}

\def\roughly#1{\mathrel{\raise.3ex\hbox{$#1$\kern-.75em\lower1ex\hbox{$\sim$}}}}
\def\barpk{{\raise.35ex\hbox{${\sss (}$}}--{\raise.35ex\hbox{${\sss )}$}}}
\def\bbarp{\hbox{$B$\kern-0.9em\raise1.4ex\hbox{\barpk}}}
\def\Kstarzero{\widehat{K^*}{\raise0.75ex\hbox{\scriptsize 0}}}
\def\Kstarplus{\widehat{K^*}^{\lower0.75ex\hbox{\tiny +}}}

\def\e{{\rm e}}
\def\o{{\rm o}}
\def\sss{\scriptscriptstyle}
\def\Re{{\rm Re}}
\def\Im{{\rm Im}}

\def\Kstarzero{\widehat{K^*}{\raise0.75ex\hbox{\scriptsize 0}}}
\def\Kstarplus{\widehat{K^*}^{\lower0.75ex\hbox{\tiny +}}}

\def\dsp{\displaystyle}

\def\adir00{{a_{\sss dir}^{00}}}

\def\B00{B^{00}}
\def\Bp0{B^{+0}}

\def\dsp{\displaystyle}

\def\gkstkpi{g_{\sss{\!K^*\!\!K\!\pi}}}
\newcommand{\tev}{\ensuremath{\mathrm{Te\kern -0.1em V}}\xspace}
\newcommand{\gev}{\ensuremath{\mathrm{Ge\kern -0.1em V}}\xspace}
\newcommand{\mev}{\ensuremath{\mathrm{Me\kern -0.1em V}}\xspace}
\newcommand{\kev}{\ensuremath{\mathrm{ke\kern -0.1em V}}\xspace}
\newcommand{\ev}{\ensuremath{\mathrm{e\kern -0.1em V}}\xspace}
\newcommand{\gevc}{\ensuremath{{\mathrm{Ge\kern -0.1em V\!/}c}}\xspace}
\newcommand{\mevc}{\ensuremath{{\mathrm{Me\kern -0.1em V\!/}c}}\xspace}
\newcommand{\gevcc}{\ensuremath{{\mathrm{Ge\kern -0.1em V\!/}c^2}}\xspace}
\newcommand{\mevcc}{\ensuremath{{\mathrm{Me\kern -0.1em V\!/}c^2}}\xspace}

\begin{document}

\title{Determination of Weak Amplitudes using Bose Symmetry
  and
Dalitz Plots}

\author{Rahul Sinha} \affiliation{The Institute of Mathematical
  Sciences, Taramani, Chennai 600113, India} 
\author{N.~G.~Deshpande}
\affiliation{Institute of Theoretical Science, University of Oregon,
  Eugene, OR 94703, USA} 
\author{Sandip Pakvasa}
\affiliation{Department of Physics and Astronomy, University of
  Hawaii, Honolulu, HI 96822, USA} 
\author{Chandradew Sharma}
\affiliation{Dept. of Physics,BITS, PILANI-K.K. Birla Goa Campus,
  Zuarinagar, Goa 403726, India}

\date{\today}
\begin{abstract}
  We present a new method using Dalitz plot and Bose symmetry of pions
  that allows the complete determination of the magnitudes and phases
  of weak decay amplitudes.  We apply the method to process like $B\to
  \kstar \pi$, with the subsequent decay of $\kstar\to K\pi$.  Our
  approach enables the additional measurement of an isospin amplitude
  without any theoretical assumption. This advance will help in
  measuring weak phase and probing for new physics beyond standard
  model with fewer assumptions.
\end{abstract}
\pacs{13.25.Hw,11.30.Ly,12.15.Hh}
\maketitle

Hadronic weak decays are important in extracting CP violating phases
and probing the effects of physics beyond the standard model.  Our
inability to calculate all hadronic effects accurately has made these
tasks challenging. While significant progress has been made in
estimating hadronic effects, one still needs to use symmetry
arguments, such as SU(3), to reduce the number of hadronic parameters
to be calculated. An alternative helpful approach is to look for
innovative methods that enable obtaining more observables, thereby
reducing the dependence on theoretical assumptions. In this letter, we
present a new method based on Dalitz plot, isospin, and Bose symmetry
that enables the measurement of extra observables and allows for a
complete determination of all the weak decay amplitudes and phases. We
present the method through the concrete example of $B\to \kstar\pi$
where the $\kstar$~\cite{f1}  decays into
$K\pi$. We show how the consequences of Bose symmetry between the two
final state pions enables one additional measurement--the direct
measurement of the 
the $A_{3/2}$ amplitude, which is crucial in determining the CP
violating phase and looking for new physics; our method is the only
one not needing any extra assumptions.


The large number of $D$ and $B$ mesons produced in heavy flavor
facilities has prompted a revival of Dalitz plot analysis approach to
study their
decays~\cite{hfag,Deshpande:2002be,Gronau:2003ep,Cheng:2005ug,
  Ciuchini:2006kv,Gaspero:2008rs,Gronau:2010kq,Bhattacharya:2010tg,
  Lorier:2010xf}. It is well known that identical bosons obey Bose
symmetry in the Dalitz plot distribution, and amplitudes must be
written in terms of isospin and spatial parts in such a way that
overall symmetry under permutation of identical particles is
obeyed. Historically, this fact has been noted in the Dalitz plot
study involving three pion decay of mesons~\cite{Zemach:1963bc}.

The four decay modes $B^0\to K^{*+} \pi^-$, $B^+\to K^{*0}\pi^+$,
$B^0\to K^{*0}\pi^0$ and $B^+\to K^{*+}\pi^0$ can be described using
isospin in a fashion analogous to the decays $B\to K\pi$. The isospin
$I=\frac{1}{2}$ initial state decays to a final state that can be
decomposed into either $I=\frac{1}{2}$ or $I=\frac{3}{2}$ via a
Hamiltonian that allows $\Delta I=0$ or $\Delta I=1$ transitions. The
transition $\Delta I=0$ results only in a single amplitude with final
state $I=\frac{1}{2}$ labeled as $B_{1/2}$, whereas the transition
with $\Delta I=1$ can results in two amplitudes with $I=\frac{1}{2}$
or $I=\frac{3}{2}$ represented as $A_{1/2}$ and $A_{3/2}$
respectively.  The isospin amplitudes $A_{1/2}$, $A_{3/2}$ and
$B_{1/2}$ are themselves defined~\cite{Lipkin:1991st} in terms of the
Hamiltonian to be: 
\begin{eqnarray}
  \label{eq:a1/2}
  A_{1/2}&=&\pm\sqrt{\tfrac{2}{3}}\langle\frac{1}{2},\pm\frac{1}{2}|{\mathscr
    H}_{\Delta I=1}|\frac{1}{2},\pm\frac{1}{2}\rangle~,\nn \\ 
  A_{3/2}&=&\sqrt{\tfrac{1}{3}}\langle\frac{3}{2},\pm\frac{1}{2}|{\mathscr
    H}_{\Delta I=1}|\frac{1}{2},\pm\frac{1}{2}\rangle~,\nn \\ 
  B_{1/2}&=&\sqrt{\tfrac{2}{3}}\langle\frac{1}{2},\pm\frac{1}{2}|{\mathscr
    H}_{\Delta I=0}|\frac{1}{2},\pm\frac{1}{2}\rangle~.
\end{eqnarray}

\begin{figure}[b]
\centering
\includegraphics*[scale=0.4]{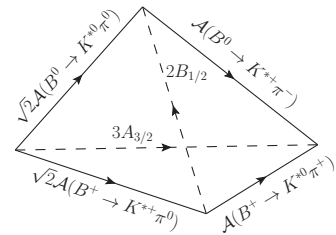}
\caption{\label{fig:quad}The four $B\to \kstar \pi$ amplitudes related
  by isospin (see Eq.~(\ref{eq:isospin})) 
. The quadrangle is fixed  once $A_{3/2}$ is
  known.}
\end{figure}

The four decays are described in terms of three isospin
amplitudes $A_{1/2}$, $A_{3/2}$ and $B_{1/2}$ as follows:
\begin{equation}
  \label{eq:isospin}
  \begin{split}
    &{\cal A}^{-+} \!=\! {\cal A}(B^0\to {\kstar}^+\pi^-)
    \!=\! A_{3/2} + A_{1/2} - B_{1/2} ,\\
    &{\cal A}^{+0} \!=\! {\cal A}(B^+\to {\kstar}^0\pi^+)
    \!=\! A_{3/2} + A_{1/2} + B_{1/2} , \\   
    &{\cal A}^{00}\!=\! \sqrt{2}\,{\cal A}(B^0\to {\kstar}^0\pi^0)
    \!=\! 2 A_{3/2} - A_{1/2} + B_{1/2} , \\   
    &{\cal A}^{0+} \!=\! \sqrt{2}\,{\cal A}(B^+\to {\kstar}^+\pi^0)
    \!=\! 2 A_{3/2} - A_{1/2} - B_{1/2}.   
  \end{split}
\end{equation}
These amplitudes satisfy the identity ${\cal A}^{00}+{\cal
  A}^{-+}={\cal A}^{+0}+{\cal A}^{0+}$ and may be represented by a
quadrilateral in the complex plane shown in Fig.~\ref{fig:quad}.
Unfortunately, much like the description of $B\to K\pi$ decays, the
four branching ratios are not enough to fix the three complex isospin
amplitudes and assumptions like SU(3) have been used to understand
these decays. We will show how Bose symmetry can be used to
obtain the isospin amplitudes directly. 

The $\kstar\pi$ final state has the distinct advantage over $K\pi$, as
it allows for a model independent measurement of all the isospin
amplitudes. To understand better how we achieve this consider for
example the decay $B^+\to K^{*0}\pi^+$, with $K^{*0}$ decaying to
$K^0\pi^0$. We hence have the final state $K^0\pi^0\pi^+$.  The two
pions, $\pi^+\pi^0$ can either have a net isospin $I_{\pi\pi}=1$ or
$I_{\pi\pi}=2$. The key point we rely on is that Bose symmetry demands
that the isospin state $I_{\pi\pi}=2$ requires that the two pions are
in spatially even state. The state $K^0\pi^0\pi^+$ with the pions
being in the even state cannot arise from final isospin 1/2 state, and
can only arise from final isospin 3/2 state. Thus isolating the state
with two pions in the spatially even state is equivalent to isolating
the $I_{\pi\pi}=2$ component, which would provide a measurement of the
isospin 3/2 amplitude of $K^0\pi^0\pi^+$ and hence the $I=3/2$
component of $\kstar\pi$, since the strong decay $\kstar\to K \pi$
conserves isospin. This provides an {\em additional observable
  $A_{3/2}$, thereby enabling the quadrangle depicted in
  Figure~\ref{fig:quad} to be completely fixed.}  The decays $B\to
K\pi\pi$ have been studied earlier~ \cite{Deshpande:2002be,
  Ciuchini:2006kv,Gronau:2010kq, Lorier:2010xf}, but the methods
proposed there do not permit determination of all the amplitudes
without additional assumptions and (or) modes being considered.


Once $A_{3/2}$ is measured the quadrangle is completely fixed. One has
a total of `eleven' observables: the four decay rates for each $B$ and
$\bar B$, $A_{3/2}$ and its conjugate mode equivalent ${\bar A}_{3/2}$
and the time dependent asymmetry relating the angle between $A(B\to
\kstar^0\pi^0)$ and $A({\bar B}\to {\bar K}^{\!*}\pi^0)$. This
provides just enough observables to solve all the `six' topological
amplitudes~\cite{Gronau:1994rj} $T$, $C$, $P$, $P_{EW}$, $P_{EW}^C$
and $A$ and their `five' relative phases purely in terms of
observables and the weak phase $\gamma(\phi_3)$ which can be measured
elsewhere. This would provide valuable information on hadronic
parameters and enable clean test of physics beyond the Standard
Model. Alternatively, one can measure the weak phase
$\gamma(\phi_3)$~\cite{Neubert:1998pt-Gronau:2000az} with fewer
assumptions about hadronic matrix elements, since we have obtained two
extra observables.

We now consider the decay $B(P)\to K(p_1) \pi(p_2)\pi(p_3)$ in the
Gottfried-Jackson frame with $B$ moving in the $\hat{z}$ axis such
that the two pions go back to back with $\pi(p_2)$ at an angle
$\theta$ with $K(p_1)$. In this frame $\vec{p_2} +\vec{p_3} =
\vec{0}$. We define $s \equiv (p_2 +p_3)^2 = (P-p_1)^2$, $t \equiv
(p_3 +p_1)^2 = (P-p_2)^2$ and $u \equiv (p_1 +p_2)^2 = (P-p_3)^2$.
$t$ and $u$ can be written as:
\begin{align}
\label{eq:t}
t &\equiv a+b \,\cos\theta~,\\
\label{eq:u}
u & \equiv a-b \,\cos\theta~,
\end{align}
where, 
\begin{align}
  \label{eq:1}
  a&= \frac{M^2 +m^2_K+ 2 m^2_{\pi}-s}{2} \\
  b&= \frac{\sqrt{s -4
    m^2_{\pi}}}{2\sqrt{s}}\lambda^{1/2}(M^2,m^2_K,s)
\end{align}
and $\lambda(M^2,m^2_K,s)=(M^4+m^4_K+s^2-2 M^2m^2_K-2 M^2 s-2 m^2_K s)$.

Let us now consider the three body final state $K\pi\pi$. Since the
final state carries two pions which respect Bose symmetry, the final
state should have an overall symmetry under isospin and space for the
two pions, i.e. isospin odd states must be odd under exchange of the
two pions and the isospin even states must be even under the exchange
of the two pions. We now construct the isospin states of
$|K\pi\pi\rangle_I$. Note, we have placed a subscript `$I$' to
indicate this is just the isospin part of the state and that the state
$|K\pi\pi\rangle$ will include the spatial dependence.  The isospin
states are obtained as follows:
\begin{eqnarray}
  |{K^0\pi^0}{\pi }^+\rangle_I &\!\!=\!\!& 
\frac{1}{\sqrt{5}}|\frac{5}{2},\frac{1}{2}\rangle_{\e}+
  \sqrt{\frac{3}{10}}|\frac{3}{2},\frac{1}{2}\rangle_{\e}
  -\frac{1}{\sqrt{6}}|\frac{3}{2},\frac{1}{2}\rangle_{\o} \nn \\
  & &\mbox{}
  -\frac{1}{\sqrt{3}}|\frac{1}{2},\frac{1}{2}\rangle_{\o},
\end{eqnarray}
where the subscript `$\rm e$' and `$\rm o$' mean that the two pions in
the state are in an `even' and `odd' state respectively. We note that
these subscripts are introduced only to take note of the isospin
symmetry of the two pions.

The complete state $|K^0\pi^0\pi^+\rangle$ resulting from B decay is
then easily written as:
\begin{eqnarray}
  \label{eq:6}
  |{K^0\pi^0}{\pi}^+\rangle&\!\!=\!\!&
  \Big(\frac{1}{\sqrt{5}}|\frac{5}{2},\frac{1}{2}\rangle_{\e}+  
  \sqrt{\frac{3}{10}}|\frac{3}{2},\frac{1}{2}\rangle_{\e}\Big)\,X\nn \\
  & &
  -\Big(\frac{1}{\sqrt{6}}|\frac{3}{2},\frac{1}{2}\rangle_{\o} 
  +\frac{1}{\sqrt{3}}|\frac{1}{2},\frac{1}{2}\rangle_{\o}\Big)\,
    Y\cos\theta,
\end{eqnarray}
where $X$ and $Y\cos\theta$ are introduced to take care of the spatial
and kinematic contributions as is seen from the discussion above 
(see Eqns.~(\ref{eq:t}) and (\ref{eq:u})). 
In general, $X$ and $Y$ can be
arbitrary even functions of $\cos\theta$. We retain the subscripts $\e$
and $\o$  merely to track the even or odd isospin
state of the two pion in the three-body final state.


We define $C_{1/2}$, $C_{3/2}$ and $D_{1/2}$ as the three-body isospin
amplitudes analogously to the two-body amplitudes $A_{1/2}$, $A_{3/2}$
and $B_{1/2}$ defined in Eq.~(\ref{eq:a1/2}). We further add a
superscript `$\rm e$' or `$\rm o$' for amplitudes arising from even or
odd isospin states respectively.  The amplitudes for the decays
$B^+\to K^0\pi^0\pi^+$ is:
\begin{multline}
  \label{eq:3a}
  A(B^+\to K^0\pi^0 \pi^+)=\frac{3}{\sqrt{10}} C_{3/2}^{\e}\,X-
  \Big[\frac{1}{\sqrt{2}} C_{3/2}^{\o} \\ 
  +\frac{1}{\sqrt{2}}
  (C_{1/2}^{\o}+ D_{1/2}^{\o} ) \Big]\,Y\cos\theta~. 
\end{multline}
The other charged $B$ decay amplitudes are:
\begin{align} 
  \label{eq:3b}
 & A(B^+\to K^+\pi^- \pi^+)= \Big[-\frac{1}{\sqrt{5}} C_{3/2}^{\e}
  +\frac{1}{\sqrt{2}}( C_{1/2}^{\e}+
  D_{1/2}^{\e} ) \Big] \nn\\ &\qquad\qquad
~~~\,X 
  +\Big[ \frac{1}{2} ( C_{1/2}^{\o}+D_{1/2}^{\o} ) -
  C_{3/2}^{\o}\Big]\,Y\cos\theta,\\
   \label{eq:3c}
 &A(B^+\to K^+\pi^0 \pi^0)\!\! =-\Big[\frac{2}{\sqrt{5}} C_{3/2}^{\e}
  +\frac{1}{\sqrt{2}}( C_{1/2}^{\e}+ D_{1/2}^{\e})\Big] X
\end{align}
\begin{multline}
  A(B^+\to K^0\pi^+ \pi^0)= \frac{3}{\sqrt{10}}
  C_{3/2}^{\e}\,X +\Big[\frac{1}{\sqrt{2}}
  C_{3/2}^{\o}\\
+ \frac{1}{\sqrt{2}} 
  (C_{1/2}^{\o}+ D_{1/2}^{\o})\Big]\,Y\cos\theta~.
\label{eq:3d}
\end{multline}
The amplitudes for the neutral $B$ decay modes can analogously be
written. We emphasize again that the amplitudes expressed in
Eq.~(\ref{eq:3a})--(\ref{eq:3d}) are explicitly Bose symmetric. One
may also note that while we have considered $B^+$ decays explicitly,
the same analysis could equally well have been done with $B^0$ decays.

We now discuss an alternate approach, where we consider the decay as a
two step process. The decays $B\to\kstar\pi$ are described by the
amplitudes given in Eq.~(\ref{eq:isospin}). The $K^{*0}$
resonance decays by strong interaction into two modes: $K^0\pi^0$ and
$K^+\pi^-$. Using isospin the states $|K^{*0}\pi^+\rangle$ and
$|{\kstar}^+ \pi^0 \rangle$ may hence be expressed in terms of the
three body finals states as,
\begin{align}
  \label{eq:2to3a}
 |\kstar^0 \pi^+ \rangle  &=  \sqrt{\frac{1}{3}} \,|[K^0 \pi^0]\pi^+
\rangle -\sqrt{\frac{2}{3}}\,|[K^+\pi^-] \pi^+ \rangle,\\
  |\kstar^+ \pi^0 \rangle &= -\sqrt{\frac{1}{3}}\,|[K^+\pi^0]
  \pi^0\rangle  +\sqrt{\frac{2}{3}} \,|[K^0\pi^+] \pi^0\rangle ~.
\end{align}
Here the square bracket denotes that the particles arise from a
$\kstar$ resonance. The matrix element for the two step decay $B^+\to
[K^0\pi^0]\pi^+$ is then given by:
\begin{equation}
\begin{split}
&\mathcal{M}(B^+\to [K^0\pi^0]\pi^+)=\frac{\gkstkpi}{\sqrt{3}}\,  
(A_{3/2}+A_{1/2}+B_{1/2})~~~~\\
&\qquad\dsp \times(P+p_3)^\mu (p_1 -p_2)^\nu\frac{\Big(-g_{\mu
      \nu} +\frac{ (p_1+p_2)_\mu 
  (p_1+p_2)_\nu}{m^2_{\kstar}}\Big)}{u-m^2_{\kstar}+im_{\kstar}\Gamma_{\kstar}}\nn
\end{split}
\end{equation}
where $\gkstkpi$ takes care of the couplings and other proportionality
terms in the expression for the amplitude.  The term $
(P+p_3).(p_1-p_2)$ is easily to be $t-s$. 
Hence,
\begin{multline}
    \label{eq:mode-u}
    \mathcal{M}(B^+\to [K^0\pi^0]\pi^+)
    =\frac{\gkstkpi}{\sqrt{3}}(A_{3/2}+A_{1/2}+B_{1/2})\\
    \times\left(\frac{s-t+c}{
        u-m^2_{\kstar}+i\,m_{\kstar}\Gamma_{\kstar}}\right).\
\end{multline}
The amplitude corresponds to a $\kstar^0$ resonance at $u=m_{\kstar}$
on the Dalitz plot. Note that the amplitude can be separated into two
parts -- the isospin amplitude and the spatial part of the amplitude
given by the large round bracket. Finally, the constant $c$ is given
by 
\begin{equation}
  \label{eq:c}
  c= \frac{(M^2-m_\pi^2)(m_K^2-m_\pi^2)}{m_{\kstar}^2}~.
\end{equation}
Similarly with an intermediate $\kstar^+$ resonance one obtains,
\begin{multline}
    \label{eq:mode-t}
    \mathcal{M}(B^+\to [K^0\pi^+]\pi^0)
    =\frac{\gkstkpi}{\sqrt{3}}(2A_{3/2}-A_{1/2}-B_{1/2})\\
    \quad\times\left(\frac{s-u+c}{
        t-m^2_{\kstar}+i\,m_{\kstar}\Gamma_{\kstar}}\right)~, 
\end{multline} 
as the amplitude corresponding to the resonance $\kstar^+$ resonance
at $t=m_{\kstar}$. 

Clearly Eqns.~(\ref{eq:mode-u}) and (\ref{eq:mode-t}) taken separately
are not of the form depicted in Eqns.~(\ref{eq:3a}) and (\ref{eq:3d})
and do not respect overall Bose symmetry as is required.  The two body
even and odd isospin amplitudes for the mode
$B^+\to\kstar^{0(+)}\pi^{+(0)}$ are given by the sum and difference of
the amplitudes for $B^+\to \kstar^0\pi^+$ and $B^+\to\kstar^+\pi^0$,
and are defined to be $A_e$ and $A_o$ respectively~\cite{f2}. We then have,
\begin{align}
  \label{eq:3}
A_e&=\frac{\gkstkpi}{\sqrt{3}}\frac{3}{2}\,A_{3/2},\\
\label{eq:4}
A_o&=\frac{\gkstkpi}{\sqrt{3}}\frac{1}{2}(-A_{3/2}+2\,A_{1/2}+2\,B_{1/2})
\end{align}
The sum of the matrix element of the two contributing modes is Bose
symmetric and may be written in an explicitly symmetric form as:
\begin{widetext}
\begin{align}
  \label{eq:5}
 \mathcal{M}(B^+\to [K \pi]\pi)&=\Bigg[A_e
\Big(\frac{s-t+c}{
  u-m^2_{\kstar}+i\,m_{\kstar}\Gamma_{\kstar}}+\frac{s-u+c}{
  t-m^2_{\kstar}+i\,m_{\kstar}\Gamma_{\kstar}}\Big)\nn \\ 
&\qquad + A_o\Big(\frac{s-t+c}{
  u-m^2_{\kstar}+i\,m_{\kstar}\Gamma_{\kstar}}-\frac{s-u+c}{
  t-m^2_{\kstar}+i\,m_{\kstar}\Gamma_{\kstar}}\Big) \Bigg]. 
\end{align}
We note that using Eqns.~(\ref{eq:3}) and (\ref{eq:4}) we recover the
sum of Eqns.~(\ref{eq:mode-u}) and (\ref{eq:mode-t}). It now follows
that:
\begin{equation}
  \label{eq:2}
  |\mathcal{M}(B^+\to [K\pi]\pi)|^2= \frac{f_1 |A_e|^2+f_2 \Re(A_e
    A_o^*)+f_3 \Im(A_eA_o^*)+f_4
    |A_o|^2}{\left(\left(m_\kstar^2-t\right)^2+ m_\kstar^2 \Gamma_\kstar ^2\right) 
    \left(\left(m_\kstar^2-u\right)^2+m_\kstar^2 \Gamma_\kstar
      ^2\right)}
\end{equation}
where the denominator can be expanded as 
\begin{equation}
  \big((m_\kstar^2-t)^2+ m_\kstar^2 \Gamma_\kstar^2\big) 
  \big((m_\kstar^2-u)^2+m_\kstar^2 \Gamma_\kstar^2\big)=
  \big((a-m_\kstar^2)^2+m_\kstar^2
  \Gamma_\kstar ^2\big)^2-2 b^2 \cos ^2\theta ((a-m_\kstar^2)^2-m_\kstar^2
  \Gamma_\kstar ^2\big)+b^4 \cos ^4\theta ,\nn
\end{equation}
and after some simplification we find,
\begin{align}
  \label{eq:7}
  f_1&=  4\,(-3a+c+Q)^2 \big((a-m_\kstar^2)^2+m_\kstar^2
  \Gamma_\kstar^2\big)+ 8 b^2 (a-m_\kstar^2)(3 a-c-Q) \cos^2\theta +
  4 b^4 \cos^4\theta \\
  f_2&=  8 b \big((3 a-c-Q)\big(m_\kstar^2
    \Gamma_\kstar^2+(m_\kstar^2-a) (-4 a+c+m_\kstar^2+Q)\big)-
  b^2(-4 a+c+m_\kstar^2+Q)\,\cos^2\theta \big)\,\cos \theta \\
  f_3&=  8\, b\, m_\kstar \Gamma_\kstar \big(-(3 a-c-Q)^2 + b^2
  \cos^2\theta\big)\,\cos\theta \\
  f_4&=  b^2 \cos^2\theta \big((-4 a+c+m_\kstar^2+Q)^2+m_\kstar^2
  \Gamma_\kstar ^2\big),
\end{align}
\end{widetext}
where $Q=M^2 +m^2_K+ 2 m^2_{\pi}$. It is obvious that the even part of
$|\mathcal{M}(B^+\to [K\pi]\pi)|^2$ can be obtained by adding to
itself the same term with $t$ and $u$ interchanged. This can be
carried out in the Dalitz plot by reflecting the data around $t=u$
line and adding it. By fitting this to
a polynomial in $\cos^2\theta$, it is straightforward to extract $|A_e|$
and thus $|A_{3/2}|$ using Eq.~(\ref{eq:3}).  

We have been able to extract the even and odd parts by symmetrization,
achieved by adding the amplitudes of the contribution from two
resonances on the Dalitz plot that are related by the exchange of two
pions.  The reader may wonder how the even and odd parts for the mode
$B^+\to [K^+\pi^-]\pi^+$ could be separated, since there exists no
resonance if $\pi^+$ and $\pi^-$ are exchanged.  Note that on the
$K^*$ resonance the two-body amplitudes $A_{1/2}$, $A_{3/2}$ and
$B_{1/2}$ are related to the three-body amplitudes $C_{1/2}^{\e,\o}$,
$C_{3/2}^{\e,\o}$ and $D_{1/2}^{\e,\o}$.  Comparing Eqns.~(\ref{eq:3}) and
(\ref{eq:4}) with Eqns.~(\ref{eq:3a}) or (\ref{eq:3d}) and since $B\to [K^+\pi^0]\pi^0$ is purely even we derive:
\begin{align}
  \label{eq:8}
  C^\e_{3/2}&=\sqrt{\frac{5}{6}}A_{3/2},~C^\e_{1/2}=-\frac{A_{1/2}}{\sqrt{3}},
  ~D^\e_{1/2}=-\frac{B_{1/2}}{\sqrt{3}},\nn\\
  C^\o_{3/2}&=\frac{A_{3/2}}{\sqrt{6}},~C^\o_{1/2}=-\sqrt{\frac{2}{3}}A_{1/2},
  ~D^\o_{1/2}=-\sqrt{\frac{2}{3}}B_{1/2}.\nn
\end{align}
Using these relations in Eq.~(\ref{eq:3b}) which is already symmetric
we find that the even and odd parts of the isospin contribution to
$B^+\to [K^+\pi^-]\pi^+$ are equal, and thus even the one surviving
pole satisfies Bose symmetry.

To summarize, we have shown how Dalitz plot, isospin, and Bose
symmetry can be used to obtain extra observables without any
theoretical assumptions. We demonstrate the usefulness of this
observation by developing the method to determine completely all the
weak decay amplitudes for $B\to \kstar \pi$. Our new approach would
provide valuable information on hadronic parameters, enable clean test
of physics beyond the Standard Model and also help in measuring the
weak phase $\gamma(\phi_3)$ all with fewer assumptions about hadronic
matrix elements. The method has several further applications in three
body decays of $D$ and $B$ mesons~\cite{upcoming}.

We thank Professors Jim Brau, Tom Browder, Alexei Garmash, Nita Sinha
and Xrexes Tata for discussions. We are also grateful to Tom Browder
and Nita Sinha for helpful suggestions.  This work is supported in
part by DOE under contract numbers DE-FG02-96ER40969ER41155 and
DE-FG02-04ER41291.


\begin{thebibliography}{99}
\bibitem{f1}
By $\kstar$ we mean $\kstar(892)$ or
  $\kstar(1430)$ and similar resonances. The detailed analysis
  presented here assumes $J^P=1^-$ $\kstar$ resonance.
\bibitem{f2}We can
  include $K^0\rho^+$ contribution, but since we are only interested
  in $A_e$, and $\rho^+$ only gives pions in the odd state, we do not
  include this contribution here. Our extraction of $A_e$ is still
  unaffected. $A_e$ does, however, receive contributions from
  finalstates such as $K^+ f_0(980)$, at fixed values of $s$. The
  contributions arising from such resonances in the $\pi\pi$ channel
  populate different parts of Dalitz plot with different $\theta$
  distribution compared with $K^*$ and can be easily excluded by
  conventional techniques. Studying the distribution in terms of $\theta$
  does not increase the uncertainty in the extraction of the
  amplitudes.
\end{thebibliography}
\end{document}